\documentclass[a4paper]{SIGRAVarticle}
\usepackage{amsmath}
\usepackage{amsfonts}
\usepackage{pictexwd,dcpic}

\newcommand{\cyl}[1]{$\Delta^*_{\varepsilon (#1)}$}

\newcommand{\ie}{\emph{i.e.}}
\newcommand{\q}{e^{\frac{2\pi}{2\pi -\varepsilon(p)}}}
\newcommand{\Rangle}{\rangle\negthinspace\rangle}

\newcommand{\bgz}{\bar{\zeta}}
\Title{From random Regge triangulations to open strings}

\Authors{M.Carfora$^{1}$, C.Dappiaggi$^{1}$}

\MoreAuthors{V.L.Gili$^{1,2}$}

\Address{$^1$Dipartimento di Fisica Nucleare e Teorica, Universit\`a degli
  Studi di Pavia (Italy) and INFN, Pavia}

\MoreAddress{$^2$ Centre for Research in String Theory, Dep. of
  Physics, Queen Mary, University of London (UK)}

\def\ShortAuthors{M.Carfora, et al.}
\def\ShortTitle{From RRT to open strings}

\begin{document}
\maketitle

\begin{abstract}
  We show how Boundary Conformal Field Theory deformation techniques
  allow for a complete characterisation of the coupling between the
  discrete geometry inherited uniformizing a random Regge
  triangulations and open string theory.
\end{abstract}

\vspace{1cm}
\emph{To appear in the Proceeding of the XVII Sigrav Conference, 4-7 September 2006, Turin}

\vspace{1.5cm}

\section{Introduction}
In the last years we witnessed a diffuse use of simplicial
techniques as powerful tools to understand the dynamical genesis of
open/closed string dualities
\cite{Dijkgraaf:2002fc,Gaiotto:2003yb,Gopakumar:2003ns,Gopakumar:2004qb,Gopakumar:2004ys,Gopakumar:2005fx,Aharony:2006th,David:2006qc}.

In this connection, one of the most interesting achievements has been
developed in the context of AdS/CFT correspondence
\cite{Gopakumar:2003ns, Gopakumar:2004qb,Gopakumar:2004ys,
  Gopakumar:2005fx,Aharony:2006th,David:2006qc} describing a special
class of $\mathcal{N}=4$ SYM free field diagrams as skeleton graphs:
from one hand, a Schwinger parametrisation of these diagrams and a
consequent change of variable in parametrization moduli space allows
to rewrite them as closed string amplitudes in $AdS$ space. Moreover,
out of the introduction of the ribbon graph baricentrically dual to
the skeleton diagram, in \cite{Gopakumar:2005fx} Gopakumar suggests to
exploit the Strebel theorem \cite{Mulase-Penkava} in order to
construct a specific punctured closed string worldsheet (and hence a
closed string vertex operators correlator) out of a given gauge theory
amplitude.

However, being a formal map between the moduli space of metric ribbon
graphs of genus $g$ with $n$ boundary components and the decorated
moduli spaces of $n$-punctured closed Riemann surfaces, namely
$\mathcal{M}_{g,n}\times\mathbb{R}_+^n$, the Strebel theorem does not
provide a dynamical description for the open-to-closed worldsheet
transition. Accordingly, we decided to look for more general settings
in which ``Strebel-like'' techniques do represent only an important
but not fully exhaustive ingredient.

To this end, we depicted a metrized Random Regge Triangulation (RRT),
which is topologically characterized by the number of its vertexes,
edges and faces, $(N_0, N_1, N_2)$, as a uniformization of an open
Riemann surface $M_\partial$ with a set of $N_0$ annuli
\cite{Carfora:2001gi,Carfora:2002rn}:
\begin{equation}
\label{eq:annuli}
  \text{\cyl{p}} \,\doteq\,
  \left\{
    \zeta(p) \,\in\, \mathbb{C}
    \,\left\vert\,
    e^{- \frac{2 \pi}{2 \pi \,-\, \varepsilon(p)}}
    \,\leq\,
    |\zeta(p)|
    \,\leq\, 1\right.
  \right\},\qquad p\,=\,1,\,\ldots,N_0 
\end{equation}
each of which is defined in the neighborhood of the $p$-th vertex of T
and it is endowed with the correspondent Euclidean cylindrical metric
$ |\phi(p)|\,\doteq\, \frac{L(p)^2}{4\pi^2}\, \vert \zeta (p) \vert^{-
  2} \vert d \zeta (p) \vert^2$. Exploiting a conformal
transformation, each $(\text{\cyl{p}},|\phi(p)|)$ can be equivalently
interpreted as a finite cylinder of circumference length $L(p)$ and
height given by $\frac{L(p)}{2 \pi - \varepsilon(p)}$. In this way, we
are trading the localised curvature degrees of freedom
associated to the parent triangulation,  specified by the
deficit angles $\varepsilon(p)$ decorating its vertexes, into modular
data associated to the new discrete surface: $\tau(p) \,=\, i
\theta(p) \,=\, i(2\pi \,-\, \varepsilon(p))$.  The decorated Riemann
surface is subsequently constructed glueing the above local
uniformizations along the pattern defined by the ribbon graph $\Gamma$
baricentrically dual to the parent triangulation.

The simplicial nature of these geometries allows to suppose that we
can implement some specific limit processes on their geometric
parameters which ``closes the holes'', hence mapping the open surface
into a closed one.  In this framework, it would be interesting to
develop the dynamical coupling between the above geometries and a
matter field theory, subsequently checking the effect of these limit
processes on the field theory couplings.  In the next sections, we
will give an answer to the first one of the above questions.

\section{Boundary Conformal Field Theory on discrete open Riemann
  surfaces: Boundary Insertion Operators}

The geometric construction we briefly sketched in the introduction
looks at the triangulated surface as getting decomposed into its
fundamental cylindrical components of finite height. These are glued
together along the pattern defined by the ribbon graph $\Gamma$, which
has been introduced as the edge refinement of the 1-skeleton
barycentrically dual to the parent triangulation. In our picture, each
cylindrical end can be interpreted as an open string connected through
its inner boundary to the ribbon graph. In this connection, the latter
acts naturally as the locus on which $N_0$, a priori independent,
Boundary Conformal Field Theories interact.  Thus, to quantize a
conformal field theory on the full $M_\partial$, it is then natural to
first describe the associated BCFT on a single cylindrical end.
Afterwards, we will describe the interaction scheme along $\Gamma$.

The fundamental prerequisite to quantize a CFT on a surface with
boundary is to have full control of the same quantum theory on the
complex plane, usually referred to as the \emph{bulk theory}. It is
defined via a suitable assignment of an Hilbert space of states
$\mathcal{H}^{(C)}$ endowed with the action of an Hamiltonian
operator, $H^{(C)}$, and of a vertex operation, \ie{} a formal map
$\Phi^{(C)}(\circ;\,\zeta,\,\bgz): \; \mathcal{H}^{(C)}
\,\rightarrow\, \text{End}\left[V [\zeta,\bgz]\right]$ associating to
each vector $\vert\phi\rangle \in \mathcal{H}^{(C)}$ a conformal field
$\phi(\zeta,\,\bgz)$ of conformal dimension $(h,\bar{h})$.  The bulk
theory is completely worked out once we know the coefficients of the
Operator Product Expansion (OPE) for all fields in the theory (for a
review on the topic, see \cite{Recknagel:1998ih,Gaberdiel:2002iw} and
references therein).  Actually, this task is tractable for most CFTs
since, among conformal fields, a preferential role is played by chiral
ones, which we will denote with $W^a(\zeta)$ and
$\overline{W}^a(\bgz)$. They are defined on the whole complex plane,
hence they can be Laurent expanded and their modes, $W^a_n$ and
$\overline{W}^a_n$, $n \in \mathbb{Z}$, generate two isomorphic and
commuting copies of the chiral algebra which defines the symmetries of
the theory, namely $\mathcal{W}$ and $\overline{\mathcal{W}}$. Beside
allowing an immediate definition of the bulk Hamiltonian, the action
of $\mathcal{W}$ and $\overline{\mathcal{W}}$ determines a diagonal
decomposition of the Hilbert space into subspaces carrying their
irreducible representations:
\begin{equation}
  \label{eq:chirdec}
  \mathcal{H}^{(C)} \,\doteq\,
  \bigoplus_{\lambda\,\overline{\lambda}}
  \mathcal{H}_\lambda
  \,\otimes\,
  \overline{\mathcal{H}}_{\overline{\lambda}},
\end{equation}
where $\lambda$ and $\overline{\lambda}$ are respectively the
$\mathcal{H}$ and $\overline{\mathcal{H}}$ highest weights.

The above data allow us to discuss the extension of such a CFT on a
given cylindrical end over $M_\partial$. To this avail, let us
remember that, microscopically, to define a CFT on a surface with
boundary means to work out which are the field values we can
\emph{consistently} assign on the boundary of the new domain. Hence,
specialising to a single \cyl{p}, the recipe we followed saw, as first
step, to look for the most generic set assignments.  In this
connection, the double nature of cylindrical ends, which can be
considered both as open string one-loop diagrams (with time flowing
around the cylinder) and tree-level diagrams for a closed string
propagating for a finite length path (with time now flowing along the
cylinder), allows to encode the possible boundary assignments into a
set of coherent boundary states defined on the inner and outer
boundaries of \cyl{p}. According to previous remarks, the expression
of such boundary states must be consistent with the algebra of field
identified by the bulk theory. On a practical ground, this is realised
requiring the absence of information flows across the boundary
components: to this end, we ask holomorphic and antiholomorphic chiral
fields to be related on it by an automorphism of the chiral algebra
\cite{DiFrancesco}.  If we conformally map \cyl{p} into an annulus, we
require on its outer boundary:
\begin{equation}
\label{eq:uguali?}
\left.\zeta(p)^{h_W} W(\zeta(p))\right|_{|\zeta(p)|=1} = \Omega\,
\left.\zeta(p)^{\bar{h}_{\overline{W}}}
\overline{W}(\bgz(p))\right|_{|\zeta(p)|=1},  
\end{equation}
$h_W$ being the conformal
weight of $W(\zeta(p))$.

The above relation has a twofold value. From one hand, radial
quantization translates it into a glueing condition for boundary
states \cite{Recknagel:1998ih}, hence allowing to select, in the above
cited set, those elements compatible with the chiral algebra of the
bulk theory. On the other side, the glueing automorphism relates on
the boundary the holomorphic and antiholomorphic chiral fields,
allowing to introduce a single one defined as:
\begin{equation}
  \label{eq:1al}
  \mathbf{W}_{\Omega} =
  \begin{cases}
    W(\zeta(p)) &
    |\zeta(p)| \leq 1 \\
    \Omega \overline{W}(\bgz(p)) &
    |\zeta(p)| > 1 \\
  \end{cases}.
\end{equation}
After the identification $\zeta^* = \bgz$, $\mathbf{W}_{\Omega}$ is an
analytic function on the full complex plane, hence it can be Laurent
expanded and its modes define a single copy of the chiral algebra
associated to the boundary conformal field theory on \cyl{p}
\cite{Cardy1,Recknagel:1998ih}. In this way, this single chiral
algebra $\mathcal{W}$ induces a decomposition of the \emph{open} CFT
Fock space $\mathcal{H}^{(O)}$ into a sum of carriers of its
irreducible representations \cite{Gaberdiel:2002iw}:
$\mathcal{H}^{(O)} \, = \, \bigoplus_\lambda \mathcal{H}_\lambda$,
being $\mathcal{H}_\lambda$ the subspace appearing in
\eqref{eq:chirdec}.

Accordingly, we can summarise the fundamental data which describe
the BCFT on a single cylindrical end \cyl{p} with:
\begin{itemize}
\item $\mathcal{Y} = \{\lambda(p)\}$, the collection of indexes
  labelling the irreducible representations of the chiral algebra
  associated to the BCFT on \cyl{p};
\item $\mathcal{A} = \{A(p)\}$, the set of possible boundary
  conditions we can assign on each boundary component, hence located
  at $|\zeta(p)|=1$ and $|\zeta(p)|=\q$ in the annuli picture. Each
  $A(p)$ includes either the glueing automorphism $\Omega_{A(p)}$,
  either a specification for all other necessary parameters. To each
  boundary condition we can associate the boundary state $\Vert
  g_{A(p)}\Rangle$.
\end{itemize}

Such characterisation of the set of boundary states representing the
admissible field assignments on the boundaries is instrumental for the
next step of our description, namely to discuss the interaction of the
pairwise adjacent BCFT copies along the pattern defined by the ribbon
graph. To this end, let us consider two adjacent cylindrical ends
\cyl{p} and \cyl{q}: they are glued to the oriented boundaries
$\partial\Gamma_p$ and $\partial\Gamma_q$ of the ribbon graph.  Let us
consider the oriented strip associated with the edge $\rho^1(p,q)$ of
the ribbon graph and its uniformized neighbourhood
$\left(U_{\rho^1(p,q)},z(p,q)\right)$. Let the boundary condition on
$\partial\Gamma_p$ and $\partial\Gamma_q$ be respectively
$A(p)\in\mathcal{A}$ and $B(q)\in \mathcal{A}$. They are specified,
among the other parameters, by a choice of the glueing automorphisms,
namely $\Omega_{A(p)}$ and $\Omega_{B(q)}$, which, according to
equation \eqref{eq:1al}, leads to the definition of the chiral fields
$\mathbf{W}_{\Omega_{A(p)}}(z(p,q))$ and
$\mathbf{W}_{\Omega_{B(q)}}(z(q,p))$.

The existence of pairwise adjacent boundary conditions does not allow
us to apply to our model the standard BCFT prescriptions which,
assuming the presence of a vacuum state not invariant under the action
of the Virasoro (translation) operator $L_{-1}^{(H)}$, allows
a boundary condition to change along a single boundary component.

\begin{figure}[!t]
  \begin{center}
    \includegraphics[width=.35\textwidth]{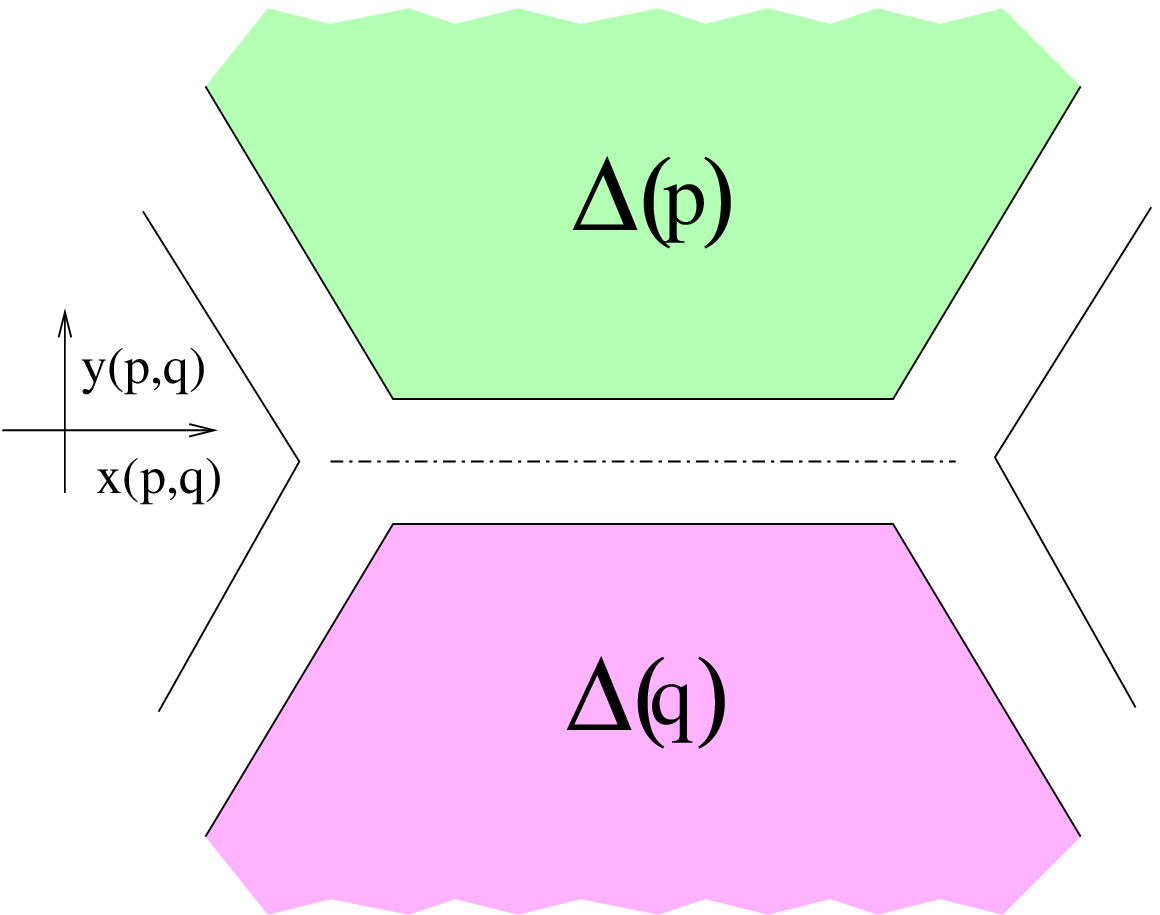}
    \quad
    \raisebox{1.5cm}{$\xrightarrow[y(p,q) \rightarrow 0]
      {\text{ADJACENCY LIMIT}}$}
    \quad
    \includegraphics[width=.3\textwidth]{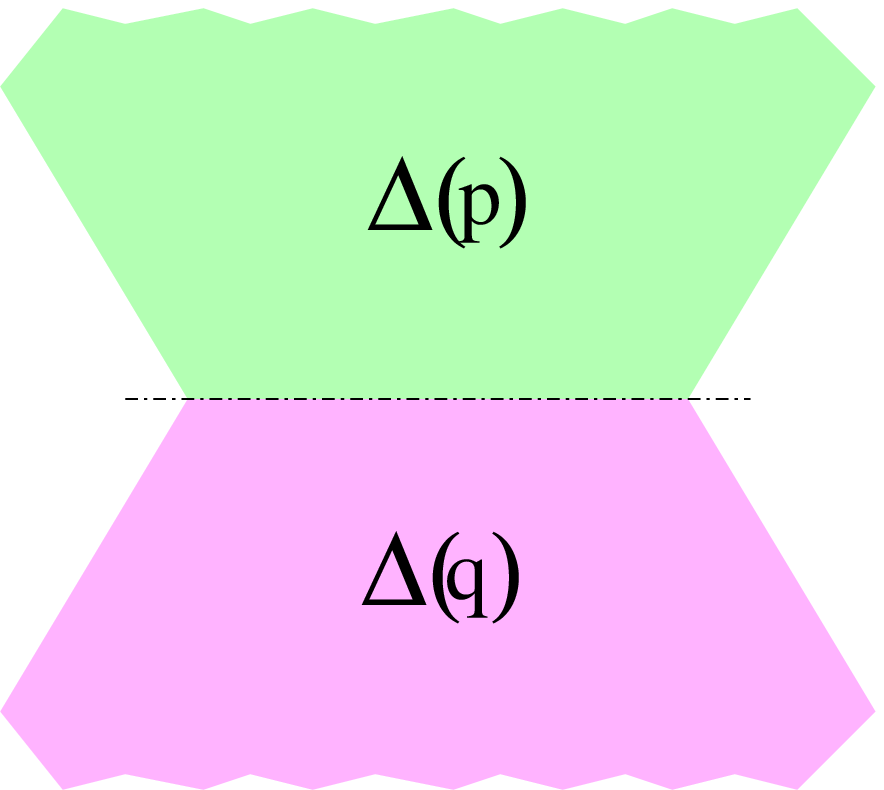}
    \caption{Shared boundaries in the adjacency limit.}
    \label{fig:BIO}
  \end{center}
\end{figure}

On the contrary, in the framework dual to a Random Regge
Triangulation, the $N_0$ cylinders are pairwise glued through a single
ribbon graph edge.  Hence, in this case, we should more properly speak
of a ``separation edge'' between two adjacent cylindrical ends (even
if, with an abuse of terminology, we will keep referring to
$\rho^1(p,q)$ as a boundary).  Furthermore, we do not have a jump
between two boundary conditions taking place at a precise point. On
the opposite, two different boundary conditions coexist in the
adjacency limit along the whole edge \cite{Carfora:2002rn}, as
depicted in fig. \ref{fig:BIO}.

%%%%%%%%%%%%%%%%%%%%%%%%%%%%%%%%%%%%%%%%%%%%%%%%%%%%%%%%%%%
%%%%%%%%%%%%%%%%%%%%%%%%%%%%%%%%%%%%%%%%%%%%%%%%%%%%%%%%%%%

Switching back to field theoretical contents, in this connection it is
no longer correct to claim the presence of a vacuum state invariant
under translations along the boundary: as a matter of fact, the shared
boundary is obtained out of two separate loops, each of them being
part of a domain where a BCFT is constructed, and all the associated
Fock space elements are invariant under translation only along the
relevant boundary loop.

Thus, in order for the geometric glueing process to be consistent with
the functional data of the theory on each cylinder, we must require
that the $N_0$ a priori independent Fock spaces blend pairwise without
breaking the conformal and the chiral symmetry of the model.

Within this framework we can implement a non symmetry-breaking glueing
of two adjacent cylindrical ends associating to such a pair a unique
copy of the chiral algebras and of the Virasoro ones. Taking into
account $z(q,p) = - z(p,q)$, we perform the glueing requiring a
condition similar to \eqref{eq:uguali?}  to hold. In this process, the
subtle point resides in the maps $\Omega$ in \eqref{eq:1al}. As a
matter of fact, we must take into account that the whole process must
relate the two glueing automorphisms $\Omega_{A(p)}$ and
$\Omega_{B(q)}$ associated to the BCFTs defined respectively on
\cyl{p} and \cyl{q}. Thus it seems natural to introduce a further
automorphism ${\Omega'}^{A(p) B(q)}$ which, in the adjacency limit
$y(p,q)=\Im\left[z(p,q)\right] \rightarrow 0$, acts along the boundary
deforming continuously the (holomorphic and antiholomorphic components
of the) bulk chiral fields in \cyl{p} into the corresponding
counterpart on \cyl{q}.  To rephrase:
\begin{equation}
\label{eq:ifr}
\mathbf{W}_{\Omega_{A(p)}} (z(p,q))|_{y(p,q) \rightarrow 0} =
{\Omega'}^{A(p) B(q)}\mathbf{W}_{\Omega_{B(q)}}(z(p,q)) |_{y(p,q)
\rightarrow 0}.
\end{equation}

In this way, we are indeed implementing a two way dynamical flow of
informations between \cyl{p} and \cyl{q}.  As a matter of fact,
\eqref{eq:ifr} provides a concrete mean to associate to each pairwise
adjacent set of conformal theories a unique chiral current out of
\eqref{eq:ifr}:
\begin{equation}
\label{eq:pq_cur}
  \mathbb{W}_{{\Omega'}_{A(p) B(q)}}\left(z(q,p)\right) =
  \begin{cases}
    \mathbf{W}_{\Omega_{A(p)}}\left(z(q,p)\right) & \text{in\;\cyl{p}
    $\cup\rho_1(p,q)$} \\
    {\Omega'}^{A(p) B(q)}\mathbf{W}_{\Omega_{B(q)}}\left(z(q,p)\right) &
    \text{in\;\cyl{q}$\cup\rho_1(q,p)$}
  \end{cases}.
\end{equation}

Expanding in Laurent series the field $\mathbb{W}_{{\Omega'}_{A(p)
    B(q)}}\left(z(q,p)\right)$ in \eqref{eq:pq_cur}, we can associate
a unique copy of the chiral algebra, namely $\mathcal{W}(p,q)$, to
each pairwise adjacent pairs of BCFTs.

We have now introduced all the main ingredients we need in order to
coherently define a full-fledged boundary conformal field theory on
the whole surface $M_\partial$.  As a matter of fact, we can associate
to each $(p,q)$ pair of BCFTs defined on adjacent cylinders along
a ribbon graph edge, a unique Hilbert space of states
$\mathcal{H}^{(p,q)}$; the latter can be determined through the action
of chiral modes of \eqref{eq:pq_cur} on a true vacuum state, whose
existence is granted per hypothesis. As usual, $\mathcal{H}^{(p,q)}$
gets decomposed into a direct sum of subspaces
$\mathcal{H}_{\lambda(p,q)}$ which are carrier of an irreducible
representation of the $\mathcal{W}(p,q)$ algebra itself. Exploiting
the state-to-field correspondence, we can associate to each highest
weight state in $\mathcal{H}_{\lambda(p,q)}$ a primary field which we
shall refer to as \emph{Boundary Insertion Operators} such that
\begin{equation}
\label{eq:bio}
\psi_{\lambda(q,p)}^{A(p) B(q)}(x(q,p)) \,=\,
\psi_{\lambda(p,q)}^{B(q) A(p)}(x(p,q)),
\end{equation}
where $x(q,p)\,=\,\Re\left[ z(q,p)\right]$.  In \eqref{eq:bio} the
notation is chosen with the following convention: $\lambda(p,q)$ is
the representation label while the decoration with indexes $A(p)$ and
$B(q)$ points out that the switch in boundary conditions actually
refers to all parameters which specify the boundary assignment.

At this stage, BIOs are purely formal objects. However, their
definition as conformal primaries allows to describe them as Chiral
Vertex Operators, hence associating them a conformal dimension related
to the highest weight of the associated Verma module.  Nonetheless, in
\cite{gili-phd} we showed that the trivalent structure of the ribbon
graph, on which BIOs are naturally defined, is sufficient to provide
all the fundamental data defining their interaction. As a matter of
fact, it allows to introduce their two point functions and OPEs as
objects which are well defined respectively on ribbon graph's edges
and vertexes.

In the next session, we will find out in which particular cases we can
characterize explicitly both their algebraic and analytic
descriptions. Moreover, we will show that, once again, the structure
of $\Gamma$ allows to fix the algebraic form of the OPEs'
coefficients.

\section{Characterisation of Boundary Insertion Operators}
\label{sec:due}

To investigate BIOs properties, let us focus on the bulk theory
identified by $D$ real embedding maps (scalar fields) $X^\alpha:\,
M_\partial \rightarrow \mathcal{T}$, $\alpha = 1, \ldots, D$, and let
us introduce the following worldsheet action on the single \cyl{p}:
\begin{equation}
  \label{eq:action}
  S
  \,=\,
  \frac{1}{4\pi}
  \int d \zeta(p) d \bgz(p)\,
  E_{\alpha \beta}(p) \left(\partial X^\alpha(p)
    \bar{\partial} \overline{X}^\beta(p)\right).
\end{equation}
In \eqref{eq:action}, the background matrix $E(p) = G(p) + B(p)$
encodes target space information by specifying the background metric
and Kalb-Ramond field components, respectively $G_{\alpha \beta}$ and
$B_{\alpha \beta}$. In particular, let us deal with flat toroidal
backgrounds, in which the $D$ directions are compactified and $E(p)$'s
components are $X$-independent.  The above configuration has an
straightforward open-string interpretation: we are dealing with $N_0$
cylindrical ends whose outer boundary can, a priori, lay on a stack of
$N$ $D$-branes. The latters, on one hand allow to decorate each
open-string with a suitable assignment of $U(N)$ Chan-Paton factors
(consequences of this from the glueing point of view are deeply
analysed in \cite{gili-phd}) while, on the other side, they act as
sources of gauge fields, whose dynamic is encoded in the previous
action (in the case of static brane and constraint field strength ) by
means of the identification $F_{\alpha \beta} = \frac{1}{4\pi}
B_{\alpha \beta}$. This obviously means that, if we consider a stack
of $D-p$-branes, the Kalb-Ramond field is constrained to have non-zero
components only along the first $p+1$ directions.

After the above premises, which ultimately show that the
characterisation of the toroidal backgrounds we will deal with
directly resides into the choice of a particular point into moduli
space of inequivalent toroidal compactifications of $D$ directions,
\ie{} \cite{Johnson:Dbranes,Giveon:1994fu}:
$$\mathcal{M}=O(D,D,\mathbb{Z})\backslash O(D,D)/[O(D)\times
O(D)],$$ let us switch back to our main aim, namely the description of
the algebraic structure and of the action of Boundary Insertion
Operators.  To this end, let us pick up those special orbits in moduli
space which are fixed under the generalised $T$-duality group
$O(D,D,\mathbb{Z})$.  These moduli values give rise to theories in
which the fundamental $[U(1)_L \times U(1)_R]^D$ current symmetry is
enanched to different symmetry groups of rank at least $D$.  In
particular, if we break $U(N)\,\rightarrow U(1)^N$, in each abelian
subsector we can choose the maximally enhanced symmetry background as
\cite{Giveon:1994fu}:
\begin{subequations}
  \begin{gather}\label{bkBG}
    G_{\alpha \beta} = \frac{1}{2}  C_{\alpha \beta}  \\
    B_{\alpha \beta} = G_{\alpha \beta} \,\,\forall\, \alpha \,>\,
    \beta,
    \quad
    B_{\alpha \beta} = -\, G_{\alpha \beta} \,\, \forall\, \alpha \,<\,
    \beta,    \quad
    B_{\alpha \alpha} = 0,\label{bkBG2}
  \end{gather}
\end{subequations}
where if
$C_{\alpha\beta},\,\alpha, \beta = 1, \ldots, D$, is the Cartan matrix
of a semisimple simply laced Lie algebra $\mathfrak{g}_D$ of total rank D. 

When dealing
with $p+1$ Neumann and $D-p-1$ Dirichlet directions, we are forced to
set $B_{m,n} = 0 \,\forall m, n = p+1,\ldots D-p-1$, hence the
maximally extended symmetry group will be:
 \begin{equation}
\label{eq:sgrup}
\mathbb{G}_D \,=\, ({G}_{p+1} \,\times{G}_{p+1}) \,\times\,  (SU(2) \times
  SU(2))^{D - p - 1}.
\end{equation}

Let us deal with a single factor ${G}_r \times G_r$ entering in
\eqref{eq:sgrup}, where ${G}_r$ is the universal covering group
generated by the simply laced Lie algebra of rank $r$,
$\mathfrak{g}_r$, whose Cartan matrix specifies the background matrix
entries. The emerging of such an extended symmetry group is one hint
of the quantum equivalence between such a theory of D free and
compactified scalar bosons and the $\hat{\mathfrak{g}}_{k=1}$-WZW
model, where $\hat{\mathfrak{g}}_{k=1}$ is the affine extension of
$\mathfrak{g}_r$ at level $k=1$.  Within this framework, the related
bulk theory can be fully characterised by the properties of the WZW
model. As a first remark, let us point out that the theory is
rational, hence the infinite serie of Verma modules can be reorganised
to write the Hilbert space of states as the direct sum of the
\underline{finitely many} irreps of the affine Lie algebra (now
playing the role of the CFT chiral algebra):
\begin{equation}
  \label{eq:gWZW_Hspace}
  \mathcal{H}^{(C)} \,=\,
  \bigoplus_{\hat{\omega}_I \in P_+^1(\hat{\mathfrak{g}})}
  \mathcal{H}_{\hat{\omega}_I}^{\hat{\mathfrak{g}}_1}
  \,\otimes\,
  \overline{\mathcal{H}}_{\hat{\omega}_I}^{\hat{\mathfrak{g}}_1}.
\end{equation}
 
Denoting (the holomorphic part of) the primary
fields associated to the highest weight state in
$\mathcal{H}_{\hat{\omega}_I}^{\hat{\mathfrak{g}}_1}$
with 
$\phi_{\hat{I}(p)}(\zeta(p))$, we immediately notice that
their components, $\phi_{[\hat{I}(p),\,m]}(\zeta(p))$, $m \,=\,
1,\,\ldots,\,\text{dim}\hat{\omega}_I$, fill the level-0 (in sense of
$L_0$ eigenvalue) subspace of
$\mathcal{H}_{\hat{\omega}_I}^{\hat{\mathfrak{g}}_1}$, which we will
denote with $\mathcal{V}_{\hat{\omega}_I}^0$. These last subspaces carry
an irreducible representation of the horizontal subalgebra of
$\hat{\mathfrak{g}}_1$:
\begin{equation}\label{eq:cla5}
  \mathbf{X}^{\hat I}_{J_0}\,:\; \mathcal{V}_{\hat{\omega}_I}^0
  \,\rightarrow\, \mathcal{V}_{\hat{\omega}_I}^0,
\end{equation}
being $J_0$ generic element of $\mathfrak{g}$ \cite{Recknagel:1998ih}.

To extend this RCFT on a surface with boundary, we can try to adopt
Cardy's construction: a set of boundary conditions that we can
consistently define on the boundaries of a cylindrical domain are
labelled exactly by the modules of the chiral algebra entering into
the Hilbert space. The correspondent boundary states are
\cite{Cardy1}:
\begin{equation}
  \label{eq:cardy_bs}
  \vert\vert \hat{\omega}_I(p)\Rangle \,=
  \sum_{\hat{\omega}_J\,\in\,P_+^1(\hat{\mathfrak{g}})}
  \frac{\mathcal{S}_{\hat{I}\,\hat{J}}}
  {\sqrt{\mathcal{S}_{\hat{0}\,\hat{J}}}}
  |\hat{\omega}_J(p)\Rangle.
\end{equation}

However, as they stand, Cardy's boundary states are necessary but not
sufficient to describe the boundary extension of our CFT. To fully
classify the plethora of boundary assignments we can coherently fix for
our model, we have to call into play the concept of BCFT deformation.

Let us remember that, given a BCFT with bulk action $S_{bulk}$,
fluctuations in the boundary condensate can move the theory away from
the Renormalisation Group fixed point, defining a new theory with
action $S = S_{bulk} + g\int{dx \psi(x)}$. If the perturbing field
$\psi(x)$ is a truly marginal boundary field, the boundary term in the
action does not break the conformal invariance; hence, the set of
coefficients $\{g\}$ entering in previous formula parametrises a set
of BCFTs differing from the unperturbed one only for a redefinition of
boundary conditions and of the related boundary states.

This is exactly what happens in our case: as a matter of
fact, let us remember that, at fixed points in toroidal
compactifications moduli space, both the set of bulk chiral currents
and the set of open string scalar states get enlarged. If we define
the new holomorphic and antiholomorphic bulk chiral currents with
$J^a(\zeta)$ and $\overline{J}^a(\bgz)$, we can represent vertex
operators associated the new open string scalars as $S^a_\lambda(u) e^{i
  \lambda_i X^i(u)} \doteq \left.\frac{1}{2}\left[J^a(\zeta) +
    \overline{J}^a(\bgz)\right] e^{i \lambda_i
    X^i(\zeta)}\right\vert_{|\zeta|=1 |\zeta| =\q}$. Hence, the
associated currents can be combined to deform the theory. The most
general perturbing term will be \cite{yegulalp,Green:1995ga}:
\begin{equation}
\label{eq:bac}
  S_B \,=\, \int d u 
  \sum_a g_a \mathbf{J}^a(\zeta)|_{|\zeta|=1}
  = \int d u 
  \left(
    \sum_{\hat{\alpha}}
    g_{\hat{\alpha}}\,e^{i\hat{\alpha}_i\,X^i}  
    +
    \sum_i g_i \partial_x X^i
  \right)\vert_{|\zeta(p)|=1}.              
\end{equation}
However, chiral marginal deformations are truly marginal ones: hence,
the deformed model will change only for a redefinition of boundary
conditions (hence boundary states and boundary field spectra).  In
this connection, the full set of boundary states can be represented as
the following rotation of the one associated to the unperturbed theory
\cite{yegulalp,Green:1995ga}:
\begin{equation}
\label{eq:newbs}
  \Vert g \Rangle \,=\, e^{i \sum\limits_{\hat\alpha}g_{\hat{\alpha}} 
  E_0^{\hat{\alpha}} \,+\,
  i \sum\limits_i g_i H^i_0} \Vert B \Rangle_{free}.
\end{equation}
where the rotation group element acting on $\Vert B \Rangle_{free}$ is
uniquely determined by the boundary action \eqref{eq:bac}.

After this rather length but necessary premise, we can came back to
our purpose, trying to describe Boundary Insertion Operators in this
background. The above characterization of boundary states is
instrumental but not completely useful to this aim. To overcome this
problem, we showed that, among the boundary states in
\eqref{eq:newbs}, we can identify Cardy's ones as those generated by a
rotation induced by an element in the centre of the universal covering
group $G_r$:
\begin{equation}
  \label{eq:ccbs}
  \Vert \hat{\omega}_I \Rangle = 
  b_{A_I}\Vert \hat{\omega}_0 \Rangle,
\end{equation}
where $b_{A_I} = e^{- 2 \pi i A_I \hat{\omega}_0 \cdot H} \in B(G_r)$
is the central element uniquely associated to algebra outer
automorphism $A_I$ which generates the fundamental weight
$\hat{\omega}_I$ out of the basic one $\hat{\omega}_0$.

Thanks to \eqref{eq:ccbs}, we introduced an alternative
parametrisation of the above set of boundary conditions as the
doublet 
\begin{equation}
  \label{eq:boundary_par}
  \left[
\Vert \hat{\omega}_I \Rangle,\, \Gamma(k)
\right], \qquad \text{with}\quad
\begin{cases}
\hat{\omega}_I & \,\in\, P_+^1(\hat{\mathfrak{g}}) \\
\Gamma(k) & \,\in\,\frac{G_r}{B(G_r)}
\end{cases}
\end{equation}
being $\Vert \hat{\omega}_I \Rangle$ a Cardy's boundary state and 
$\Gamma(k)\in\frac{G_r}{B(G_r)}$ such that:
\begin{equation}
\label{eq:poly_bs}
  \Vert g(k) \Rangle
  \,=\,
  \Gamma(k) \, \Vert \hat{\omega}_J(k) \Rangle.
\end{equation}

Moreover, we showed that the above parametrisation is not only a
formal datum, but it allows to re-describe the model as a deformation
of the $\hat{\mathfrak{g}}_{k=1}$-WZW model described ``\'a la Cardy''
by means of the boundary term $S_\Gamma = \int du \Gamma_a J^a(u)$
such that we can define its coefficients through a suitable immersion
of $\Gamma(k)$ into the universal covering group $G_r$:
\begin{equation}
  \label{eq:immer}
  \sigma:\; \Gamma \in \frac{G_r}{B(G_r)} \hookrightarrow
              e^{i\Gamma_a J^a_0} \in G_r
\end{equation}
 
This description allowed firstly to characterize completely the glueing
automorphism $\Omega_{[\hat{J}_2,\Gamma_2](q)\,[\hat{J}_1,\Gamma_1](p)}$
entering in \eqref{eq:pq_cur} by means of:
\begin{itemize}
\item the fusion coefficients
  $\mathcal{N}_{\hat{J}(p)\,\hat{I}(p,q)}^{\hat{J}(q)}$ of the WZW
  model,
\item a deformation induced by the rotation $\Gamma(p,q) =
  \Gamma_2(q)\Gamma_1(p)^{-1}$.
\end{itemize}

Moreover, we gathered the following expression for Boundary Insertion
Operators' components in the rational limit of the conformal theory:
\begin{equation}
  \label{eq:final_bio}
  \psi^{[\hat{J}_2,\,\Gamma_2](q)\,[\hat{J}_1,\,\Gamma_1](p)}
  _{[\hat{I},\,m](p,q)}
  \,=\,
  \sum_{n=0}^{\text{dim}|\hat{I}|}  
R^{\hat{I}(p,q)}_{m\,n(p,q)}(\Gamma_2{\Gamma_1}^{-1}) \,
  \psi_{[\hat{I},\,n](p,q)}^{\hat{J}_2(q)\,\hat{J}_1(p)},
\end{equation}
where $\psi_{\hat{J}(p,q)}^{\hat{J}(p)\,\hat{J}(q)} (x(p,q)) \,=\,
\mathcal{N}_{\hat{J}(p)\,\hat{I}(p,q)}^{\hat{J}(q)}
\,\psi_{\hat{I}(p,q)}(x(p,q))$ and where $R^{\hat{I}(p,q)}_{m\,n(p,q)}=
\exp\left[\frac{i}{2}X^{\hat{I}(p,q)}\right]_{m\,n(p,q)}$ being
$X^{\hat{I}}$ the operator introduced in \eqref{eq:cla5}.

\section{Discussions and conclusions}
Boundary Insertion Operators, defined as in equation
\eqref{eq:final_bio}, have a beautiful feature: the boundary
perturbation induced by $\Gamma(p,q)$ does not affect their algebra,
which hence is completely fixed in terms of the fusion rules of the
WZW-model. This is indeed a check that our prescription for the
$(p,q)$ glueing automorphism and its subsequent action on BIOs is
consistent. As a matter of fact all deformations we have introduced
are actually truly marginal ones and, thus, they must not break the
chiral symmetry defined by $\hat{\mathfrak{g}}$. Hence, the expansion
coefficients of the product between $\psi^{[\hat{J}_1,\,\Gamma_1](p)\,
  [\hat{J}_3,\,\Gamma_3](r)}_{\hat{I}_1(r,p)} (\omega_r)$ and
$\psi^{[\hat{J}_3,\,\Gamma_3](r)\,
  [\hat{J}_2,\,\Gamma_2](q)}_{\hat{I}_2(q,r)} (\omega_q)$ in terms of
$\psi^{[\hat{J}_1,\,\Gamma_1](p)\,
  [\hat{J}_2,\,\Gamma_2](q)}_{\hat{I}_3(q,p)} (\omega_p)$ can be
written uniquely as
$\mathcal{C}_{\hat{I}_1(r,p)\,\hat{I}_2(q,r)\,\hat{I}_3(q,p)}^{\hat{J}_1(p)\,\hat{J}_3(r)\,\hat{J}_2(q)}$.

\begin{figure}[!t]
  \centering
  \includegraphics[width=\textwidth]{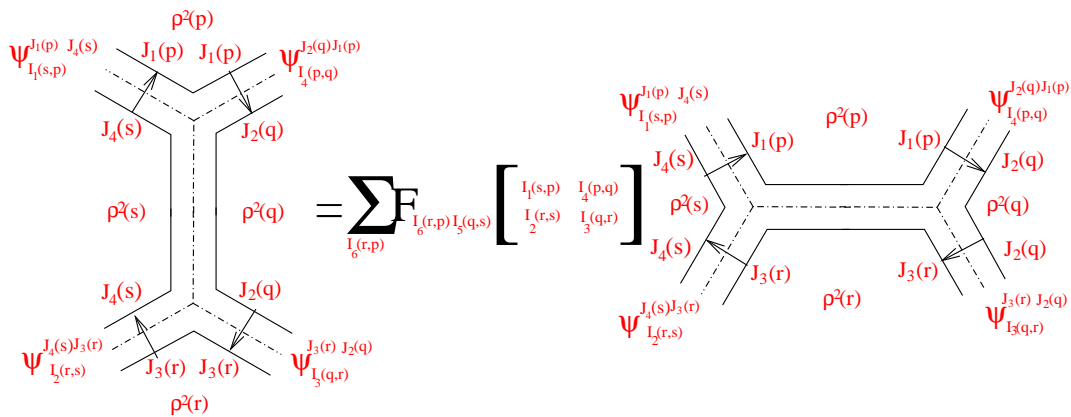}
  \caption{Four-points function crossing symmetry.}
  \label{fig:BIO_4points}
\end{figure}

The final piece of the puzzle we are trying to make up comes directly
from the study of the four-point function among BIOs. This is
naturally defined in the neighbourhood of four adjacent cylindrical
ends, hence its computation calls into play the variable connectivity
of random Regge triangulations, which produces the same feature on the
ribbon graph.  Hence, if we notice that the two factorisations through
which we can compute the four-point function are strictly related to
the two ways we can fix the connectivity among the surrounding
polytopes (see figure \ref{fig:BIO_4points}), we can write an identity
formally equivalent to the usual Pentagonal Identity for BCFT
\cite{Runkel:1998pm,Behrend:1999bn,Felder:1999ka}. This allows to
identify BIO's OPE coefficients describing interactions in the
neighbourhood of the $(p,q,s)$ vertex of the ribbon graph with the
$\hat{\mathfrak{g}}$-WZW model fusion matrices with the following
entries assignments\cite{Alvarez-Gaume:1988vr}:
\begin{equation}
  \label{eq:OPE_fusion}
  \mathcal{C}^{\hat{J}_1(p)\,\hat{J}_2(q)\,\hat{J}_3(s)}
  _{\hat{I}_1(q,p)\,\hat{I}_2(s,q)\,\hat{I}_3(s,p)}
  \,=\,
  F_{\hat{J}_2(q)\,\hat{I}_3(s,p)}
  \begin{bmatrix}
    \hat{J}_1(p)   &  \hat{J}_3(s)   \\
    \hat{I}_1(q,p) &  \hat{I}_2(s,q)
  \end{bmatrix}.
\end{equation}

This naturally completes the programme we outlined at the beginning of
this paper. The algorithm we presented has a twofold value. From one
hand, we have been able to provide an explicit expression for the
formal rules describing the interplay among BCFTs on different
cylinders whenever we consider toroidal compactifications for the
target space of the bosonic scalar field. On the other side, the open
string interpretation we gave at the beginning of section
\ref{sec:due} provides a natural way to colour the ribbon graph
(defined as the edge refinement of the 1-skeleton barycentrically dual
to a random Regge triangulation), with labels proper of the chosen
gauge group. This ultimately leads to the constructions of a genuine
't Hooft diagram and to the definition of new kinematical background
in which we investigate dynamical processes which are characteristic
in open/closed string dualities.

\bibliographystyle{amsplain}

\end{document}